\documentclass[10pt,tightenlines,showpacs,aps,prl,twocolumn,amsmath,amssymb,amsfonts,floatfix]{revtex4}
\usepackage{amsxtra}
\usepackage{graphicx}
\begin{document}
\title{Full Counting Statistics of Spin Currents}
\author{Antonio \surname{Di Lorenzo}}\author{Yuli V. \surname{Nazarov}}
\affiliation{Department of Nanoscience, Faculty of Applied Sciences, 
Delft University of Technology, 2628 CJ Delft The Netherlands} 
\pacs{05.60.Gg, 72.25.Ba}
\begin{abstract}
We discuss how to detect fluctuating spin currents
and derive full counting statistics of electron spin transfers.
It is interesting to consider several detectors in series that 
simultaneously monitor different components of the spins transferred. 
We have found that in general the statistics of the measurement outcomes 
cannot be explained with the projection postulate 
and essentially depends on the quantum dynamics of the detectors.
\end{abstract}
\maketitle
The detection and statistics of quantum fluctuations attracts increasingly the  
attention of the physical community. 
The Full Counting Statistics (FCS)~\cite{Levitov93,Levitov96} 
of quantum electron transport provides all possible information
about fluctuations of electric current in mesoscopic systems. 
The FCS has been evaluated for charge 
transport between superconducting~\cite{Belzig01} 
and superconducting-normal leads~\cite{Belzig02} 
by nonequilibrium Green function 
method~\cite{Nazarov99}.  
An approach to FCS of a general variable presented in~\cite{Kindermann02}
allows one to resolve possible inconsistencies that concern the
quantum measurement problem by explicitly incorporating the dynamics of a 
detector.

There is a strong current interest to new ways to manipulate, control
and measure electron spins in solid state. This defines the field 
of spintronics~\cite{Awschalom02}  that aims 
to gain the same control over \emph{spin} as one 
currently has over \emph{charge}. 
This provides a motivation to study the FCS of spin current. 
This statistics is of fundamental interest since different
components of spin do not commute, this makes the problem
of corresponding quantum measurement especially relevant.

In this Letter,  
we propose a realization of a spin current detector. 
We derive the FCS of a spin current, originating from a flow of unpolarized 
particles, 
that is measured by such detector(s). 
We focus on the cases where two and three components of the spin 
current are detected simultaneously by two or three detectors in series. 
Na\"{i}vely, one could try to describe the FCS of such a combined 
measurement by applying the Projection Postulate
after the measurement by each detector.
We explicitly demonstrate that 
for the case of  three detectors 
the FCS  cannot be explained in such a straightforward way
and depends on quantum dynamics of the central detector. 

We consider two-terminal electric circuit where electrons are transferred
between the terminals through a contact. The theory of FCS is elaborated
in detail for the case when this contact can be described in the 
Landauer-B\"{u}ttiker scattering framework.
In fact, our results do not depend on the type of the contact provided
it does not polarize the spin of electrons transferred, nor such polarization
occurs in the terminals. Thus, all of our results can be applied as well to 
neutron sources. 
Since the electrons carry spin, 
the charge transfer between the terminals should be accompanied
by spin transfer although there is no average spin current between the 
terminals. Therefore, there are fluctuations of spin current. How to measure
them? 

\begin{figure}[h!]
\includegraphics[width=0.4\textwidth]{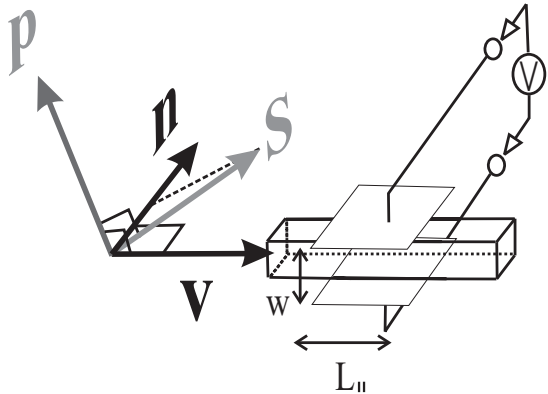}
\caption{\label{fig:detector}
The proposed spin current detector. An electron with velocity $\mathbf{v}$ 
and spin $\mathbf{S}$ induces a voltage drop in a capacitor. The electric 
field $\mathbf{E}$ inside the capacitor produces an Aharonov-Casher phase shift 
on the electrons.}
\end{figure}

This can be probably done in many ways, for instance, by exploiting
the spin-valve effect~\cite{Tsymbal01}. In the present Letter, we concentrate
on a different setup proposed and used 
in~\cite{Cimmino89} to detect Aharonov-Casher effect~\cite{Aharonov84} 
for neutrons. This setup exploits the fact that 
a moving magnetic dipole generates an electric one~\cite{Costa67}. 
To measure this, one encloses  
the two-dimensional current lead between the plates of a 
capacitor as shown in Fig.~\ref{fig:detector}. Each spin moving with
velocity $\mathbf{v}$ produces a dipole moment 
$\mathbf{d} = \frac{g}{2}\mu_B(\frac{\mathbf{v}}{c}\times\mathbf{S})$,
$g$ being the gyromagnetic factor, $\mu_B$ Bohr's magneton, $c$ the speed 
of light, and 
electron spin $\mathbf{S}$ is measured in units of $\hbar/2$.
This moment induces a voltage drop $V$ 
between the plates, which is the detector read-out, 
the variable being measured. 
Since the interaction between the dipole moment and
electric field $\mathbf{E}$ in the capacitor is 
$H_{int}=-\mathbf{E}\cdot\mathbf{d}$, the read-out signal is 
proportional to spin current
in the lead $\mathbf J$, 
$V=\lambda \mathbf{n}\cdot\mathbf{J}$, 
$\mathbf{n}$ being the unit
vector perpendicular to the direction of the current flow
and parallel to the plates of the capacitor, 
$\lambda$ being a proportionality coefficient. The concrete expression 
for the latter, $\lambda = 
\frac{g}{2}\frac{L_\shortparallel}{w}\frac{\mu_B}{C c}$,
depends on the capacitance $C$ and geometrical dimensions:
the length of its plates in the direction 
of the current $L_\shortparallel$, and the distance between the plates $w$. 

The variable canonically conjugated to the read-out is the
charge $Q$ in the capacitor, and the expression for the
interaction in terms of $Q$
contains the same proportionality coefficient $\lambda$, 
$H_{int}=-\lambda Q \mathbf{n}\cdot\mathbf{J}$. 
Our choice of the detection setup is motivated by the fact
that this detector does not disrupt electron transfers through
the contact and only gives a minimal feedback:
the electrons passing the capacitor in the direction of current
acquire Aharonov-Casher phase shift. This
phase shift depends on spin and is given by $\Phi_{AC}= \lambda Q \mathbf{n}\!\cdot\!\mathbf{S}/\hbar$.
This is similar to the detection scheme 
presented in~\cite{Levitov96} for charges transferred.
A fundamental complication in comparison with the charge FCS 
is that in our case the phase shift depends on spin, so that even the
minimal feedback may cause the rotation of spin of the electron that
passes the capacitor. 
  
It is important to require that charge and spin current
are the same in the contact and in the detector.
To assure that spin current conserves,
neither the length of the detector $L_\shortparallel$
nor the distance from detector to contact
should exceed the spin relaxation length.
In addition to this, there should be no spin or charge accumulation
between the contact and the detector. This is always true for 
low-frequency fluctuations of charge or spin current. 

\begin{figure}[htb!]
\includegraphics[width=8cm]{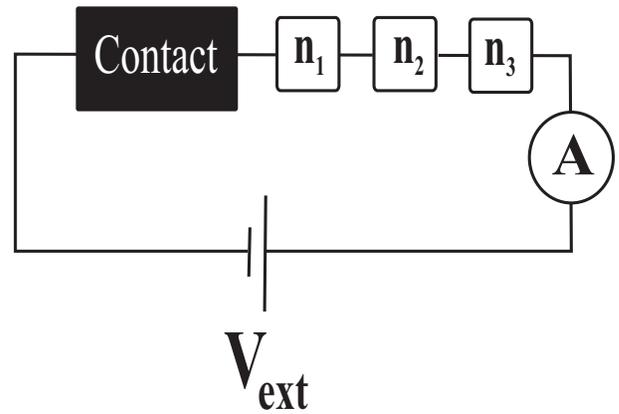}
\caption{\label{fig:setup}The setup considered, in the case of three spin 
detectors and one charge detector.}
\end{figure}

We will see that the most interesting setup includes several 
detectors in series as presented in Fig.~\ref{fig:setup}.
It may also include a charge detector. 
The spin detectors can have arbitrary polarization vectors $\mathbf{n}_a$.
This can be achieved by turning the current lead and the capacitor plates 
in different directions.  
The contact is  biased by a voltage source $V_{ext}$. Since we assume that
the resistance brought by detectors is much smaller than that of the contact,
the whole voltage drops at the contact and the contact works as a fluctuating
source of (spin) current measured by (spin) detectors.

Let us  compare the detection scheme proposed with 
the one used in Stern-Gerlach experiment~\cite{Gerlach22a}.  
In this classical experiment,
an unpolarized beam of spin $1/2$ atoms was split into two 
sub-beams corresponding to two different projections of spin 
onto an axis set by magnetic field. The intensities of the beams
are then detected, and at this stage 
the wave function collapse is said to happen: 
the spin projection of each atom detected is with certainty 
$\pm \hbar/2$ depending on the beam.  
If one would force these atoms
to pass the subsequent Stern-Gerlach detectors, the readings 
of these detectors can be predicted from this fact.

In our detection scheme, 
a single detector does the same as a Stern-Gerlach one:
it measures the difference of numbers of particles passed with 
spin ``up'' and ``down'' with respect to the polarization vector $\mathbf{n}$. 
What can we say about the readings of
subsequent detectors? 
A straightforward assumption would be that 
the second detector, like in Stern-Gerlach experiment,
measures the statistical mixture of the states with the certain
projection on the polarization vector of the first detector 
(the wave function is "collapsed" after the first measurement). 
We will refer to this assumption as to Projection Postulate (PP).
  
We will explicitly show that it is not what happens in our detection 
scheme if there are at least three detectors: the FCS of readings 
differs form the one predicted with using PP. The difference
shows up in second and fourth order cumulants of spin currents, and the fact
that the wave function is not collapsed can be just simply observed in
this way.

We adopt the fully quantum definition of FCS that has been put forward
in~\cite{Kindermann02}. We consider a number of spin current detectors
in series, labelled by index $a$ that increases from the contact to the 
terminal (Fig. \ref{fig:setup}), next to a charge detector (ammeter).
We consider the couplings between the detector degrees
of freedom and the system, $\hat H_{int}=
-\frac{\hbar}{e} \Hat\chi \hat I 
-\sum_{a} \hbar\Hat{\gamma}_a \mathbf{n}_a\!\cdot\!\mathbf{\Hat{J}}$. 
Here $\hat \gamma_{a}\equiv \lambda_a \Hat{Q}_a/\hbar$ 
are proportional to operators of charge in the 
capacitors of the corresponding spin current detectors, 
and $\hat\chi$ is the degree of freedom of the charge detector.
The FCS is defined as a kernel that relates 
the initial density matrix of
all detectors to the final one, that one after the measurement,
\begin{align}
\rho_f(\chi^+,\gamma^+_{a}; \chi^-,\gamma_{a}^-)  = & \nonumber \\
e^{-\tau F(\chi^+ -\chi^-,\gamma^+_{a},\gamma^+_{a})}&
\rho_{in}(\chi^+,\gamma^+_{a}; \chi^-,\gamma^-_{a})
\label{FCS_definition}
\end{align} 

The most general result we report here is that 
this FCS can be directly expressed in terms of
\emph{charge} counting statistics $F_c$, provided neither the contact  
nor the terminals polarize the electrons transferred. 
In this case,
\begin{equation}
	F(\chi,\{\gamma^+_{a}\},\{\gamma^-_{a}\}) 
=\frac{1}{2} \sum_{\pm} F_{c}(\chi \pm \alpha)
\label{main_relation}
\end{equation}
where $e^{\pm i\alpha}$ are the eigenvalues 
of the $2 \times 2$ matrix
\begin{equation}
\prod^{\rightarrow}_a
e^{i\gamma^+_{a}\mathbf{n}_a\cdot \boldsymbol{\sigma}}
\prod^{\leftarrow}_a
e^{-i\gamma^-_{a}\mathbf{n}_a\cdot \boldsymbol{\sigma}} \;.
\label{matrix}
\end{equation}
In this matrix, $\boldsymbol{\sigma}$ is a pseudovector of $2\times2$
Pauli matrices, and  $\prod\limits^{\rightarrow}_a (\prod\limits
^{\leftarrow}_a)$
denotes the product 
ordered with decreasing (increasing) $a$.

We arrive to the relation (\ref{main_relation}) by extending the
scattering theory of FCS. By virtue of (spin) current conservation,
the coupling terms $\hat H_{int}$ can be gauged away and ascribed to
the electron Green functions of the terminal~\cite{Levitov96,Belzig01}.
While for charge counting statistics
this gauge transform  is just a multiplication by a phase factor $e^{i\chi}$
the gauge transform generated by a spin detector $a$
involves a unitary matrix in spin space, 
$e^{i \gamma_{a} \mathbf{n}_{a}\cdot\mathbf{\sigma}}$.
The total transform is thus the product of these matrices nested 
in a proper order.
We notice that for the unpolarizing circuit this transform just commutes with 
the Hamiltonian. This enables one to derive the relation (\ref{main_relation}). 

For the case of 
one or two detectors in series, 
the eigenvalues $e^{\pm i\alpha}$
are not affected by the order of matrix multiplication in (\ref{matrix})
and depend on differences of spin counting fields 
$\gamma_a \equiv \gamma^+_{a} -\gamma^-_{a}$ only.
This implies that the FCS definition (\ref{FCS_definition}) can be readily
interpreted in classical terms: it is a generating function 
for probability distribution of a certain number of spin counts 
$S_a$
in each detector,
\begin{equation}
P(\{S_a\}) = \int \prod_{a} d \gamma_a 
e^{-\tau F(0,\{\gamma_a\})} e^{- i\sum_a S_a \gamma_{a}}
\label{probability} 
\end{equation}
For a single detector, the spin FCS is very simple: 
it corresponds to independent transfer of 
two sorts of electrons, with spins "up" and "down"
with respect to quantization axis. The (higher-order) cumulants of the 
 spin (charge) transferred
are given by (higher-order) derivative of $F$ with respect to $\gamma_a$($\chi$),
at $\chi=\gamma_a=0$. From this and relation (\ref{main_relation}) we conclude that all odd cumulants of spin current
are 0 while all even cumulants equal to even cumulants 
of the charge transferred.

It is interesting to note that in the case considered one 
can provide a "reasonable alternative" to the consistent
quantum mechanical derivation. One can evaluate FCS assuming 
that the probability to measure a certain spin count
in the detector $m$ only depends on the count of the immediately 
preceding detector $m-1$. 
This corresponds to the PP: after each measurement, 
the wave function collapses to one of the eigenstates loosing
memory about the previous evolution.
We calculate conditional probability for two detectors and then 
nest these probabilities to obtain that the probability distribution
for spin counts is given by Eq.~(\ref{probability}), with $F$ determined by  
Eq.~(\ref{main_relation}) 
with a {\it different} parameter $\alpha$, 
given by
\begin{align}\label{PPresult}
	&\cos{\alpha_{\raisebox{-1pt}{$_{PP}$}}}\!\!=\!\!
	\sum_{\mu_a=\pm 1}  
e^{i\!\sum_a\!\mu_a \gamma_a} 
\frac{1}{2}\!\prod_{a=1}^{K-1} 
\frac{1\!+\!\mu_{a} \mu_{a+1} \mathbf{n}_a\!\cdot\!\mathbf{n}_{a+1}}{2}. 
\end{align}
We note that $\alpha_{PP}$ depends only on the differences 
$\gamma_a=\gamma_a^+-\gamma_a^-$. 
Eq.~(\ref{PPresult}) facilitates the comparison of the 
results of two approaches and allows us to pinpoint
the quantum  mechanical features missed in PP analysis.  

We stress that for the case of 
one or two spin detectors, $\alpha=\alpha_{PP}$,
and the approaches give precisely the same result.
Essentially, the result for two detectors can be understood
in terms of the probability to have the same or opposite spin counts 
for one electron that passes both detectors.
The probability of having the same counts is simply 
$P_{12}=(1+\mathbf{n}_1\!\cdot\!\mathbf{n}_{2})/{2}$.
For second order cumulants --- noises --- one obtains
$\langle\!\langle S^2_1\rangle\!\rangle  
=\langle\!\langle S^2_2\rangle\!\rangle=\langle\!\langle N^2\rangle\!\rangle
$, $\langle\!\langle S_1 S_2\rangle\!\rangle = \mathbf{n}_1\!\cdot\!\mathbf{n}_{2}\langle\!\langle N^2\rangle\!\rangle$,
$\langle\!\langle N^2\rangle\!\rangle$ being the charge noise, i.e.~the second 
cumulant of the number of transferred particles.
The detector feedback is 
irrelevant, so it is possible to measure two spin components 
within the setup studied.

Let us now consider three spin detectors with arbitrary $\mathbf{n}_{1,2,3}.$
In this case, the FCS defined by Eq.~(\ref{main_relation}) 
cannot be immediately
interpreted in terms of probability distribution Eq.~(\ref{probability}). 
To illustrate
this explicitly, let us consider the change of density matrix upon 
one electron passing all detectors. In $\gamma_a$-representation it is given
by $\rho_f(\gamma^+_{a}; \gamma_{a}^-)  = 
\cos{\alpha}~
\rho_{in}(\gamma^+_{a}; \gamma^-_{a})$, where $\alpha$ in the case of
three detectors assumes the following form: 
\begin{align}\nonumber
&\cos\alpha =\cos{\alpha_{PP}} -
\sin{\gamma}_3 \sin{\gamma}_1 \left[\right.
\\ \nonumber
& \left.\cos\Gamma_2 \left( 
{\mathbf n}_1\!\cdot\!{\mathbf n}_3 -
({\mathbf n}_1\!\cdot\!{\mathbf n}_2) ({\mathbf n}_2\!\cdot\!{\mathbf n}_3) 
\right)
\!+\!\sin{\Gamma_2} ({\mathbf n}_1 \times
{\mathbf n}_2)\!\cdot\!{\mathbf n}_3
\right]
, \nonumber \\
&\cos\alpha_{PP} = \cos{\gamma}_1 \cos{\gamma}_2 \cos{\gamma}_3 \nonumber \\
&-\sin{\gamma}_1 \sin{\gamma}_2  \cos{\gamma}_3 {\mathbf n}_1\!
\cdot\!{\mathbf n}_2 
-\sin{\gamma}_2 \sin{\gamma}_3 
\cos{\gamma}_1 
{\mathbf n}_2\!\cdot\!{\mathbf n}_3 +\nonumber \\
&- \sin{\gamma}_3 \sin{\gamma}_1 \cos{\gamma}_2 
({\mathbf n}_1\!\cdot\!{\mathbf n}_2) 
({\mathbf n}_2\!\cdot\!{\mathbf n}_3)
. 
\label{avsapp}
\end{align}
where 
$\Gamma_2 \equiv \gamma^+_2 +\gamma^-_2$.
Thus the 
multiplication with $\cos\alpha$ corresponds to multiplication
with several $\exp(\pm i\gamma)$, $\exp(\pm i\Gamma_2)$ 
and adding the results with some weights.
Let us assume that initial density
matrix corresponds to the state with certain number of counts in each detector.
Since the ``number of counts'' representation is obtained from 
the $\gamma_a$-representation
by Fourier transforming with respect to $\gamma^{\pm}_a$, multiplication 
with $\exp(\pm i\gamma_a)$ transforms diagonal elements of density matrix
into diagonal ones; $\rho(S_a,S_a) \rightarrow \rho(S_a\pm 1, S_a \pm 1)$,
and the state with a well-defined number of counts remains such.
However, the multiplication with $\exp(\pm i\Gamma_2)$ produces non-diagonal
elements from diagonal ones; 
$\rho(S_2,S_2) \rightarrow \rho(S_2\pm 1, S_2 \mp 1)$.
One readily sees from Eq.~(\ref{avsapp}) that this is disregarded if one
applies PP. This seems OK since 
non-diagonal elements do not contribute to probabilities. 
However, if another electron passes the second detector,
the non-diagonal elements can be again transformed into diagonal ones 
and do contribute to the probability distribution of the counts.     
Thus, the second detector disturbs the correlation of read-outs in 
the first and third detector. The actual FCS will depend on the 
dynamics of the second detector since its feedback is unavoidable.
This feedback is eventually an Aharonov-Casher effect: the electrons
passing the second detector acquire phase shift $\pm\Gamma_2$, this corresponds
to rotation of their spin by angle $\Gamma_2$ about the axis $\mathbf{n}_2$.
Theoretically, one could prepare the second detector in a given state
and then observe the dependence of FCS on this state. However, 
in the present Letter we would like to describe a 
more realistic situation where no special preparation takes place
and the degree of freedom of the second detector 
just fluctuates following its own (dissipative) dynamics. We illustrate
the effect with a simple model of such dynamics: 
$\Gamma_2$ exhibits time-dependent
Gaussian fluctuations described by the following action
\begin{equation}
	{\mathcal S}_{det} =\frac{1}{2}\int dt\left((\dot\Gamma_2(t))^2\tau_c +
\frac{(\Gamma_2(t) - \Gamma_0)^2}{ 4 \tau_c 
\langle\!\langle \Gamma^2 \rangle\!\rangle^2} \right)
\end{equation} 
so that it fluctuates around the averaged value $\Gamma_0$ 
with the variance 
$\langle\!\langle \Gamma^2 \rangle\!\rangle$ 
and typical 
correlation time $\tau_c \langle\!\langle \Gamma^2 \rangle\!\rangle$.
The generating function of the actual FCS results from the averaging of the
Eq.~(\ref{main_relation}) over fluctuations of $\Gamma_2$,
\begin{equation}
{\mathcal Z}(\{ \gamma_a\}) = \int {\mathcal D} \Gamma(t) e^{ -{\mathcal S}_{det}
- \int_0^\tau \! F(\alpha( \{ \gamma_a\},\; \Gamma_2(t)))dt } 
\end{equation}
This allows us to compute the cumulants of spin counts and compare them
with PP expressions given by Eqs.~(\ref{probability}), (\ref{avsapp}).
The difference is in principle noticeable for the second order cumulants 
--- noise correlations in the first and in the third detectors,  
\begin{align}
&\langle\!\langle S_1 S_3\rangle\!\rangle \!=\! 
\langle\!\langle N^2\rangle\!\rangle 
\left[
C + 
\left(
\tilde{\mathbf{n}}_1\cdot \mathbf{n}_3- 
C
\right)
e^{-\langle\!\langle \Gamma^2\rangle\!\rangle/2}
\right], 
\end{align}
where 
$C \equiv (\mathbf{n}_1\cdot\mathbf{n}_2)(\mathbf{n}_2\cdot\mathbf{n}_3)$, 
$\tilde{\mathbf{n}}_1$ is the vector $\mathbf{n}_1$ rotated about 
$\mathbf{n}_2$ by the angle $\Gamma_0$,
and the first term presents 
the PP result. 

The second term, as expected, has a typical signature of interference effects:
it is suppressed exponentially if the variance of the corresponding 
Aharonov-Casher phase $\langle\!\langle\Gamma^2\rangle\!\rangle \gg 1$. Since 
$\Phi_{AC}$ is inversely proportional to $\hbar$, this is the 
classical limit. In this limit, the result coincides with the PP.

However, if one goes to fourth order cumulants --- correlations of noises ---
one observes big deviations from PP even in the classical limit. 
We obtain that
\begin{align}\label{QMpredictions}
\langle\!\langle S_1^2 S_3^2\rangle\!\rangle &\!=\! 
\langle\!\langle S_1^2 S_3^2\rangle\!\rangle_{\raisebox{-1mm}{$_{\!\!PP}$}}
+ \\ \nonumber 
&+\frac{2}{3} 
A 
\left[\langle\!\langle N^4\rangle\!\rangle -\langle\!\langle N^2\rangle\!\rangle\right]
+16\frac{\tau_c}{\tau}
A
\langle\!\langle N^2\rangle\!\rangle^2 ,
\end{align}
where \mbox{$A\equiv \left[1-(\mathbf{n}_1\!\cdot\!\mathbf{n}_2)^2\right] 
\left[1-(\mathbf{n}_2\!\cdot\!\mathbf{n}_3)^2\right]
$,} 
and the PP result is expressed in terms of charge cumulants as 
\begin{equation*}
\langle\!\langle S_1^2 S_3^2\rangle\!\rangle_{\raisebox{-1mm}{$_{\!PP}$}}\!\!=
(1+2C^2)\langle\!\langle N^4\rangle\!\rangle 
+ 2 (1-C^2) \langle\!\langle N^2\rangle\!\rangle .
\end{equation*} 
This deviation results from correlations of $\Gamma_2$ at time scale $\tau_c$.
To estimate the result, we notice that the charge cumulants are of the order
of $\tau/\tau_{el}$, $\tau_{el}$ being the average time between electron
transfers. It is easy to fulfill the condition 
$\tau_{el} \ll \tau_{c} \ll \tau$, and in this case 
$\langle\!\langle S_1^2 S_3^2\rangle\!\rangle$ is much bigger than PP result.   

In conclusion, we have discussed the full counting
statistics of spin currents measured by the detectors
that provide the minimum back-action to the passing particles. 
This FCS can be evaluated quite generally for unpolarized currents.
We compare the actual FCS with the predictions of Projection Postulate
and find an agreement for one and two detectors. However,
the FCS of three detectors in series displays both explicit dependence
on dynamics of the second detector and significant deviations from
PP predictions. These deviations can be observed in second and
fourth order cumulants of spin currents. 

We acknowledge the financial support provided through the European
Community's Research Training Networks Programme under contract
HPRN-CT-2002-00302, Spintronics. 
We are grateful to R. Fazio and A. Romito for interesting discussions.

\end{document}